\documentstyle[stwol]{article}
\begin{document}
\onecolumn
\begin{titlepage}

\begin{flushright}
  CERN--TH/96--287 \\
         hep-ph/9610351\\
\end{flushright}
       \vspace{8mm}
\title{\Large Optimized-Variational Quark Mass Expansion\\
and Dynamical Chiral Symmetry Breakdown~\footnote{Talk given at the
28th International Conference on High Energy Physics (ICHEP96),
25-31 July 1996, Warsaw, Poland, to appear in the proceedings.
Contribution supported in part by EC contract CHRX--CT94--0579.}    }
\author{\large Jean-Lo\"{\i}c Kneur~\footnote{On leave
from Laboratoire de Physique Math\'ematique et Th\'eorique,
U.R.A. 768 du C.N.R.S.,  F-34095 Montpellier Cedex 5, France.}}
\date{}

\address{\large \em CERN, Theoretical Physics Division \\
CH-1211 Geneva 23 Switzerland }
\vspace{1cm}
\maketitle
\begin{abstract}
\setlength{\baselineskip}{15pt}
{\large A recently proposed variational mass expansion approach
to
dynamical chiral
symmetry breakdown is reviewed. We briefly explain
how a specific integral ansatz over
the analytically continued, arbitrary Lagrangian mass parameter,
resums the usual variational mass expansion
(``delta-expansion".)
The construction is then generalized to obtain
non-perturbative
expressions for the order parameters of the $SU(n_f)_L \times SU(n_f)_R$
breakdown ($n_f =$ 2 or 3), in QCD.
Emphasis is put on some general aspects as well as
possible limitations of
this approach.}
\end{abstract}
\vspace{6cm}
\begin{flushleft}
CERN--TH/96--287 \\
October 1996
\end{flushleft}

\end{titlepage}

\vfill\eject
 \clearpage


\def\Journal#1#2#3#4{{#1} {\bf #2}, #3 (#4)}

\def\NCA{\em Nuovo Cimento}
\def\NIM{\em Nucl. Instrum. Methods}
\def\NIMA{{\em Nucl. Instrum. Methods} A}
\def\NPB{{\em Nucl. Phys.} B}
\def\PLB{{\em Phys. Lett.}  B}
\def\PRL{\em Phys. Rev. Lett.}
\def\PRD{{\em Phys. Rev.} D}
\def\ZPC{{\em Z. Phys.} C}

\def\st{\scriptstyle}
\def\sst{\scriptscriptstyle}
\def\mco{\multicolumn}
\def\epp{\epsilon^{\prime}}
\def\vep{\varepsilon}
\def\ra{\rightarrow}
\def\ppg{\pi^+\pi^-\gamma}
\def\vp{{\bf p}}
\def\ko{K^0}
\def\kb{\bar{K^0}}
\def\al{\alpha}
\def\ab{\bar{\alpha}}
\def\be{\begin{equation}}
\def\ee{\end{equation}}
\def\bea{\begin{eqnarray}}
\def\eea{\end{eqnarray}}
\def\CPbar{\hbox{{\rm CP}\hskip-1.80em{/}}}
\def\ds{\displaystyle}
\newcommand \dsty \displaystyle
\newcommand{\ggt}{\tilde\Gamma}
\newcommand{\mo}{m_{0}}
\newcommand{\go}{g_{0}}
\newcommand{\beq}{\begin{equation}}
\newcommand{\eeq}{\end{equation}}
\newcommand{\Lam}{\Lambda_{\overline{MS}}}
\newcommand{\eps}{\epsilon}
\newcommand{\MS}{\overline{MS}}
\newcommand{\qq}{\langle \bar q q \rangle}
\newcommand{\nn}{\nonumber}

\bibliographystyle{unsrt}    




\title{\large Optimized-Variational Quark Mass Expansion\\ 
and Dynamical Chiral Symmetry Breakdown}
\author{ Jean-Lo\"{\i}c Kneur~$^\star$}
\address{CERN,
\em Theoretical Physics Division, 
CH-1211 Geneva 23, Switzerland}

\twocolumn[\maketitle\abstracts{
A recently proposed variational mass expansion approach
to 
dynamical chiral
symmetry breakdown is reviewed. We briefly explain 
how a specific integral ansatz over 
the analytically continued, arbitrary Lagrangian mass parameter, 
resums the usual variational mass expansion
(``delta-expansion".) 
The construction is then generalized to obtain 
non-perturbative 
expressions for the order parameters of the $SU(n_f)_L \times SU(n_f)_R$
breakdown ($n_f =$ 2 or 3), in QCD.
Emphasis is put on some general aspects as well as
possible limitations of 
this approach. 
}]

\section{Introduction}
\setcounter{footnote}{0}
To calculate 
the low-energy properties of the QCD spectrum from ``first principle"
is clearly a desirable task, but obviously out of the reach of ordinary
perturbation theory, due to the infrared growth of the coupling and the 
non-perturbative
relevant dynamics. 
Chiral perturbation theory~\cite{GaLeut} 
gives a consistent low-energy effective description, but   
in terms of a set of parameters to be fixed from the data, 
whose precise connection with the basic QCD coupling and quark mass 
parameters is far
from being resolved at present. Of particular relevance to the low-energy
QCD dynamics are the order parameters of the $SU(n_f)_L 
\times SU(n_f)_R$
($n_f=$ 2,3) chiral symmetry breakdown (CSB), 
such as the $\qq$ condensate or the
pion decay constant $F_\pi$, typically.
Partially 
related to the previous aim, it has however been
known for a long time that, at least in simplified field-theoretic models, 
definite information on specific non-perturbative quantities 
may be inferred from particular resummation properties~\cite{renormalons}
 and/or 
appropriately modified perturbation 
series~\cite{delta}. 

Recently~\cite{qcd1,qcd2}
we have investigated a new approach
to explore {\it how far}
the basic QCD Lagrangian can provide
non-zero
quark condensates, pion decay constant, as well as 
{\em dynamical} quark
masses,
in the limit of vanishing Lagrangian (current) quark masses. 
The starting point is very similar to
the idea developed long ago 
in refs.~\cite{delta}, 
where 
the convergence of ordinary perturbation
was shown to be improvable by a variational procedure in which
the separation of the
action into ``free" and ``interaction" parts
is made to depend on some set of auxiliary parameters~\footnote{
In one-dimensional field theories this optimized perturbation
theory (``delta-expansion") 
has been shown~\cite{converg} to lead to a rigorously convergent
series of approximations, even in strong coupling cases.}.

\noindent An essential novelty
is that our construction
can reach
{\it infinite} order~\cite{gn1,gn2} of
such a variational-perturbative expansion, therefore presumably optimal, 
provided it converges. It also gives a 
consistent treatment of the renormalization, reconciling
the variational expansion with the inherent infinities
of most field theory of dimension higher than 1. 
\section{Mass Gap in the Gross-Neveu model}
We shall first illustrate the basic ingredients of our construction
with a determination~\cite{gn1,gn2} of 
the mass gap in the $O(2N)$ Gross-Neveu (GN) model.

From renormalization group (RG) resummation properties, one can infer
the following form of a {\em bare}, RG-invariant resummed mass
($D=2+\eps$):
\beq
m_F(\mo) = \mo\;(1- 4\pi b_0\: g^2_0
\:\ggt m^{\eps}_F )^{-\frac{\gamma_0}{2b_0}  }\;,
\label{mf1}
\eeq
where $m_0$, $g_0$ are the bare mass and coupling, 
$\ggt \equiv \Gamma[-\eps/2]/(4\pi)^{1+\eps /2}$, and
$b_0$, $\gamma_0$ are the one-loop RG-coefficients of the running
coupling and mass, respectively (we use a normalization 
such that $\beta(g) = -b_0 g^3 -b_1 g^5 -\cdots$,
$\gamma_m(g) = \gamma_0 g^2 +\gamma_1 g^4
+\cdots$; where in the $O(2N)$ model $b_0 =(N-1)/(2\pi)$,
$\gamma_0 =(N-1/2)/\pi$, etc~\cite{Gracey}.)
As easily shown, eq.~(\ref{mf1}) gives, after
renormalization, the exact mass gap result in the $N \to \infty$,
$m \to 0$ limit:
$m_F = \Lam $, where $\Lam \equiv \bar\mu \:e^{-1/(2b_0 
g^2(\bar\mu))}$ is the basic scale in the $\MS$ scheme. 
Now, in the more complicated case of arbitrary $N$, 
according to the above-mentioned 
variational principle, 
we expect to obtain
a series of approximants to the mass gap, by optimizing with respect
to the {\em arbitrary} Lagrangian mass, at successive orders of the
variational $x$-series, formally defined  
by substituting everywhere in the bare Lagrangian:
\beq m_0 \to m_0\; (1-x); ~~~~g^2_0 \to g^2_0\; x,
\label{substitution}
\eeq
and in particular  
for example in expression (\ref{mf1}). However, 
in order to get
a finite and non-trivial result (i.e. $m_F \neq 0$) one has
to resum 
the resulting $x$-series by using  
a specific contour integration~\cite{gn2} of (\ref{mf1}) 
over the (analytically continued)
$x$ parameter.   
Moreover, the one-loop RG-resummed expression (\ref{mf1}), which 
includes only the leading dependence in $m$, 
has to be generalized to
include both higher order RG-dependence upon 
$m$, {\em and} non-logarithmic
purely perturbative corrections as well, in order to get a more realistic
mass gap in the arbitrary $N$ case.
In this way we end up with the ansatz: 
\bea
{ M^P_2 (m^{''})\over \Lam}
 = {2^{-C} m''\over{2 i \pi}} \oint dv {e^{\;v} 
\over{F^A(v) [C + F(v)]^B}}  \nn \\
\times {\left(1 +{{\cal M}_{1}\over{F(v)}}
+{{\cal M}_{2}\over{F^2(v)}}+\cdots \right)},
\label{contour7}
\eea
where the contour is around the negative real axis, 
\beq
F(v) \equiv \ln [m''v] -A \; \ln F -(B-C)\; \ln [C +F],
\label{Fdef}
\eeq
with $A =\gamma_1/(2 b_1)$, $B =\gamma_0/(2 b_0)-\gamma_1/(2 b_1)$,
$C = b_1/(2b^2_0)$; 
$\Lam$ is the (RG-invariant) scale 
at two-loop order;
for convenience we introduced 
the scale-invariant, arbitrary (dimensionless) ``mass" parameter,
redefined in terms of $m$ as 
\beq
m''\equiv  \ds{\left(\frac{m(\bar\mu)}{ \Lam}\right) \;
2^{C}\;[2b_0 \bar g^2]^{-\gamma_0/(2b_0)}
\;\left[1+\frac{b_1}{b_0}\bar g^2\right]^B}
\; .
\label{msec2def}
\eeq
As mentioned, $F(1)$ in the integrand of (\ref{contour7}) 
by construction 
resums the leading and next-to-leading logarithmic
dependence in $m(\bar\mu)$ to all orders~\cite{qcd2}.
The two-loop non-logarithmic perturbative coefficients 
${\cal M}_{1}$, ${\cal M}_{2}$, given explicitly in ref.~\cite{gn2},  
connect the {\em pole} mass with 
the running mass $m(\bar\mu)$. 
 
We have shown~\cite{gn2} that 
expression (\ref{contour7}) does  
resum 
the $x$-series~\footnote{$v$ in (\ref{contour7}) 
is related to the original
expansion parameter $x$ as $x = 1-v/q$, $q$
being the order of the $x$-expansion.} generated from the substitution
(\ref{substitution}).
Moreover it is possible to choose a 
renormalization scheme (RS) such that $b_i = \gamma_i =0$ for
$i \geq 2$.
In that sense, eq.~(\ref{contour7}) resums
the full RG dependence in $x$ and $m''$. 
In contrast, the purely perturbative (non-logarithmic)
information, contained in ${\cal M}_{1}$, ${\cal M}_{2}$, is
limited to the two-loop order. 
This is where the variational principle
and optimization play their role, whereby we hope to obtain a sensible
approximation to the exact mass gap.
Accordingly, we may now look for extrema of 
expression (\ref{contour7}) with respect
to $m''$, using standard contour integration techniques.
Observe in fact 
that, were we in a simplified theory where 
${\cal M}_{1} = {\cal M}_{2} = \cdots = 0$ 
(as incidentally is the case in the large-$N$ limit), 
(\ref{contour7}) would have a very simple
behaviour near its optimum (at $m'' \to 0$), 
giving a simple pole 
with residue $M_2 = (2C)^{-C}\;\Lam $.
Now, in the arbitrary $N$ case where ${\cal M}_1$, ${\cal M}_2$,... 
cannot be neglected, one can construct a set of
approximants of (\ref{contour7}) in the $m'' \to 0$ limit, 
with some variants of Pad\'e approximants (PA). The results~\cite{gn2}
of two different optimal
fifth-order PA 
are compared in the table below
with exact results for the mass gap 
of the $O(2N)$ model  
(obtained alternatively from the Bethe ansatz in ref.~\cite{FNW}).
As illustrated, an error of ${\cal O}$(5\%) or less, depending
on $N$, 
can be obtained~\footnote{The results given in the table 
were obtained by further optimizing with respect to an arbitrary scale
parameter, introduced from $\bar\mu \to a\:\bar\mu $. Results from
PA of order lower than the (optimal) order 5 give a larger error.}. 
\begin{table}[htb] 
\begin{tabular}{||l||l|l|l||}  \hline
  & Exact  &  Pad\'e 1 &  Pad\'e 2 \\
 $N$    & $m_F/\Lam $ & (error)  & (error)  \\ \hline 
 2 & 1.8604   &  1.758 (5.5\%) &    1.875 (0.8\%) \\ \hline
3 & 1.4819 
&  1.475 (0.5\%) & 1.486 (0.3\%)  
\\ \hline
5 & 1.2367 
& 1.284 (3.7\%) & 1.265 (2.3\%)
\\ \hline
\end{tabular}
\label{tabl2}
\end{table}
\vspace{-5pt}
\section{Dynamical quark masses}
Most of the previous construction (for arbitrary $N$)
may be formally extended to the QCD case. In particular, 
expression (\ref{contour7}),  
with the appropriate change to 
QCD values
of the $b_i$ and $\gamma_i$ RG coefficients, provides a 
dynamical mass ansatz~\cite{qcd1,qcd2}
 as a function of the quark flavours $n_f$
(the latter entering both $\Lam$ and the $b_i$, $\gamma_i$ expressions).
It is important to note 
that expression (\ref{contour7}) for arbitrary $N$ in the
GN model uses exactly the {\em same} 
amount of perturbative and RG information
as is at our disposal at present for a QCD quark mass: namely,
the {\em exact} two-loop RG-resummed 
plus purely perturbative ${\cal M}_1$, ${\cal M}_2$~\cite{Broadhurst} 
dependence.
Since our construction only relies on RG-properties (and 
analytic continuation
techniques), passing from 2 to 4 dimensions 
is not expected to cause major changes, at least naively~\footnote{
From that point of view,
there are {\em no} differences between QCD and 
the GN model (for arbitrary $N$),
as compared e.g. with 
other two-dimensional 
(or a fortiori, one-dimensional) 
theories where particularly simple features are 
sometimes due to their super-renormalizable properties. 
Note also that
a mass gap in the $O(2N)$ GN model 
does not contradict 
the Coleman theorem~\cite{Coleman} on the non-breaking of the
continuous chiral symmetry in two dimensions. 
This in a sense illustrates that
the applicability of our method is more general than the physics of
CSB.}.

Actually, one complication {\em does} occur: 
as a more careful examination of 
relation (\ref{Fdef}) indicates, there are  
extra branch cuts 
in the $v$ plane, with ${\rm Re}[v_{cut}] > 0$
for the relevant case of $n_f =$ 2 or 3 in QCD.
This prevents using
the expansion near the origin, 
which would lead to
ambiguities of ${\cal O}(\exp(+{\rm const}./m''))$
for $m'' \to 0$. 
The origin of those singularities is rather
similar to the renormalon ones~\cite{renormalons}, since they appear  
in a resummed expression relating a ``reference" scale
$M_{dyn} \simeq \Lam $ to an infrared scale $m'' \simeq 0$.
However, in the present
construction
it is possible~\cite{qcd2} to move those extra cuts to 
the safe location ${\rm Re}[v^{'}_{cut}] \leq 0$, by noting that
the actual position of those cuts
depend on the RS, via
$\gamma_1$.  
Performing thus a second-order 
perturbative RS change in $m(\mu)$, $g(\mu)$, 
which changes $\gamma_1 \to \gamma^{'}_1$,
and looking for a flat optimum (plateau) of
an appropriately constructed PA, with respect to the
remnant RS arbitrariness~\footnote{See
ref.~\cite{qcd2} for details.}, 
we obtain:
\beq
M^{Pad\acute{e}}_{opt}(m''\to 0) \simeq 2.97\;\Lam(2)\;
\label{Mnum}
\eeq
for $n_f=2$, and a similar result for $n_f =3$.
\section{Order parameters of CSB: $F_\pi$ and $\qq$}
The above dynamical quark mass, although it has some meaning 
as regards spontaneous CSB in QCD, hardly has a direct physical
intepretation, e.g. 
as a pole in the S-matrix, due to confinement.
In other words, it is not a properly defined order parameter. 
It is however possible to generalize the mass ansatz (\ref{contour7}) 
to obtain  
a determination of the ratios $F_\pi/\Lam$ and $\qq(\mu)/\Lam^3$.
The latter gauge-invariant quantities are 
unambiguous order parameters, i.e. $F_\pi \neq 0$ {\em or} $\qq \neq 0$ 
indicate spontaneous CSB. 
The appropriate generalization~\footnote{A non-trivial
point in this generalization is the necessary additional
(additive) renormalization of the axial vector-axial vector 
two-point correlator,
whose first-order expansion term with respect to external momentum 
defines $F_\pi$~\cite{GaLeut}. 
For a fixed RS, this unambiguously leads to (\ref{Fpiansatz}).}
of (\ref{contour7}) for $F_\pi$ is~\cite{qcd2}
\bea
& \ds{{F^2_\pi \over{\Lam^2}} = (2b_0)\;
{2^{-2 C} (m'')^2\over{2 i \pi}} \oint {dv\over v}\; v^2 {e^{\: v}}}
\; \times \nn \\
& \ds{ \frac{1}{F^{\;2 A-1} [C + F]^{\;2 B}} 
 \; \delta_{\pi }
 \left(1 +{\alpha_{\pi}\over{F}}+{\beta_{\pi}
\over{F^2}}
\right) }
\label{Fpiansatz}
\eea
in terms of the very same 
$F(v)$ defined in eq.~(\ref{Fdef}) (therefore leading to the same
extra cut locations as in the mass case), and where
$\delta_\pi$, $\alpha_\pi$ and $\beta_\pi$
are fixed by matching the perturbative $\MS$
expansion, known to 3-loop order~\cite{abdel,Avdeev}. 
A numerical optimization with respect to the RS-dependence
of an appropriate PA, along the 
same line as the mass case, gives e.g for $n_f =2$: 
\beq
F^{Pad\acute{e}}_{\pi ,opt}(m'' \to 0)
 \simeq 0.55\;\Lam(2)\;.
\label{Fpinum}
\eeq
Concerning $\qq$, an ansatz similar 
to (\ref{Fpiansatz}) can be derived (with 
coefficients $\delta$, $\alpha$, $\beta$ 
specific to $\qq$ and obvious changes in the 
$m''$, $F$ and $v$ powers),
but for the RG-invariant combination $m \qq$ {\em only},
since our construction  
only makes sense for RG-invariant quantities.
As it turns out, the 
limit $m \to 0$ simply gives $m\qq \to 0$. To extract an estimate of 
the (scale-dependent) condensate $\qq(\mu)$ in our framework is only  
possible by introducing~\cite{qcd2} 
an explicit symmetry-breaking
quark mass $m_{exp}$ 
(independent from $m$), 
and expanding the $m\qq$ ansatz to first order in $m_{exp}$. This gives
e.g. for $n_f =2$: 
\beq
\qq (\bar\mu = 1\;\mbox{GeV}) \simeq 0.52 \; \Lam(2)\;.
\label{qqnum}
\eeq 
Comparing the results (\ref{Mnum}), (\ref{Fpinum}) and (\ref{qqnum})
gives a fairly small value of the quark condensate (and a fairly
high value of the dynamical mass), as compared to other non-perturbative
methods~\cite{sumrules}. Although small values of $\qq$ are 
not experimentally excluded
at present~\cite{Stern}, it is also clear that our relatively crude 
approximation deserves more refinements for more realistic QCD predictions.
\section{Summary and conclusions}
The variational expansion in arbitrary $m$, developed
in the GN model~\cite{gn2}, can thus be formally
extended to the QCD case.
This gives non-trivial relationships between
$\Lam$ and the dynamical masses and order parameters, $F_\pi$
and $\qq$.

Let us conclude with some remarks on the possible 
generalizations (and limitations) of this  
approach. In principle, it may be applied to any (renormalizable or
super-renormalizable) field theory, where the
massless fermion (or scalar) limit 
is of interest, and most probably
the basic idea could be applied to parameters other than mass.
Now, what is at any rate limitative, is the 
relatively poor knowledge
of the purely perturbative part of the expansion (only known to 
two-loop order in most
realistic field theories). Our ansatzes (\ref{contour7}),
(\ref{Fpiansatz}) would be exact if there 
were no such corrections, i.e. if the dependence on $m$ was 
entirely dictated
from RG properties. Therefore, our final numerical results
crucially depend on the optimization~\footnote{ 
For instance, PA results for (\ref{contour7}), (\ref{Fpiansatz}) 
are substantially different~\cite{qcd2} 
in the ({\em unoptimized}) $\MS$ scheme.}. Apart from a few
models where the series is known to large orders
(as in the
anharmonic oscillator~\cite{delta,bgn}, 
or in the GN model for $N \to \infty$),
we can hardly compare successive orders of the variational-optimized
expansion to estimate, even qualitatively, its convergence properties.
Invoking the ``principle of 
minimal sensitivity"~\cite{delta}, although physically 
motivated, in a sense 
artificially 
forces the series to converge, with no guarantees that it is 
toward the right result. 
\section*{Acknowledgements} 
Some of these results were obtained in 
works~\cite{gn2,qcd1} done in collaboration
with
C. Arvanitis, F. Geniet, M. Iacomi and A. Neveu.
This contribution is supported in part by EC contract CHRX--CT94--0579. 

\vspace{5pt}
\noindent $^\star$ On leave
from LPM,
URA 768 CNRS, F-34095 Montpellier, France.
\end{document}